\documentstyle[12pt]{article}
\begin{document}
\centerline {To appear in the Nova collection of papers dedicated to the
Lorentz symmetry,}
\centerline {V. Dvoeglazov, Editor, Nova Publisher}

\vspace*{2cm}
\begin{center}
{\large \bf UNIVERSALITY OF THE ISOSPECIAL RELATIVITY
FOR THE INVARIANT DESCRIPTION OF ARBITRARY SPEEDS OF LIGHT}
\end{center}

\vskip 0.5cm
\centerline{{\bf Ruggero Maria Santilli}}
\centerline{Institute for Basic Research}
\centerline{P.O.Box 1577, Palm Harbor, FL 34682, U.S.A.}
\centerline{E-address ibr@gte.net; Web Site http://home1.gte.net/ibr}

\begin{abstract}
\noindent We review the recent experimental evidence on arbitrary
values of the
speed of light within physical media (interior dynamical problems); we
recall
that classical (operator) theories with speeds
of light different than that in vacuum are
noncanonically (nonunitarily) related to the special relativity,
thus losing the original invariant character with consequential rather
serious
problems of physical consistency; we show that the isominkowskian
geometry, the
isopoincar\'e symmetry and the isospecial relativity provide a
formulation of arbitrary speeds of light 
within physical media as invariant as that of the
special relativity for the propagation of light  in vacuum; we prove
that the isospecial
relativity is directly universal, in
the sense of holding for all possible
spacetime models with a symmetric metric directly in the frame of the
observer; we outline the broader genotopic and hyperstructural
formulations which are
directly universal for single- and multiple-valued irreversible systems,
respectively, with a generally nonsymmetric metric; we review the
anti-isomorphic isodual formulations for the classical and
operator representation of antimatter and the prediction of the
new "isodual photon"; we finally point out a few
representative connections between the invariant isotopic formulations
and a rather large variety of generalized theories
existing in the literature.
\end{abstract}

\vskip 0.5cm

\noindent {\bf 1. Local character of the speed of light}.
\vskip 0.5cm

\noindent One of the most majestic
achievements of this century for mathematical
beauty, axiomatic  consistency and experimental verifications has been
the  {\it  special  relativity}  (SR) [1].

Nevertheless, science is a discipline that will never admit "final
theories".
No matter how valid a given theory may appear at a given point in
time, its generalization for broader physical conditions should
be expected as inevitable.

Along the latter lines, the SR has been challenged during the recent
decades
on various grounds. That primarily addressed in this note is the
experimental evidence
of the {\it local} character of the speed of light c, i.e., its
dependence on
the local spacetime coordinates x, the local density $\mu$, the
frequency $\omega$, and any other needed variable, according to the
familiar law
\begin{equation}
c = c_o / n(x, \mu, \omega, ...),
\label{eq:one}\end{equation}
where n is the index of refraction, with the propagation in vacuum
at the constant speed $c_o$ admitted as a particular case for $n = 1$.

In fact, speeds $c = c_o/n < c_o, n > 1$ (also called
{\it subluminal speeds}), are known to exist in our Newtonian
environment since the discovery of the
refraction of light in the transition
from air to water. Lesser known is the fact that one of the
first invariance studies of speeds $c < c_o$
was done by Lorentz [2a]
(see the related mention in
Pauli's book [2b]).

In short, incontrovertible experimental evidence establishes
that electromagnetic waves have all locally different speeds
when propagating in water, glass, oil, or other media.

Speeds $c = c_o / n > c_o, n < 1$ (also called {\it superluminal
speeds}),
have been recently measured by A. Enders and G. Nimtz [2c]
in the tunneling of photons between certain guides. Speeds
$c = c_o/n > c_o,$ have also been measured for large masses in
certain astrophysical explosions [2e-2g]. The literature on
the experimental measurements of superluminal speeds of electromagnetic
waves
is nowadays rather vast and it is herein assumed as known
(see also review [2d], the recent data [2h] and the
comprehensive review [2i].

The lack of exact
validity of the SR for all these different local speeds
of light $c < c_o$ and $c > c_o$ is then beyond scientific
or otherwise credible doubt, the only scientific issue being
the identification of a more adequate theory.

Note that the old hopes of regaining the exact validity of the SR
by reducing light to photons scattering
among molecules are no longer
viable because: 1) the reduction, e.g., of a {\it classical}
electromagnetic wave in our atmosphere
with one meter wavelength to photons in {\it second  quantization}
cannot evidently eliminate
the need for a classical representation prior to quantum forms acquiring
credibility;
2) the reduction does not permit quantitative studies of
superluminal speeds; and 3) the reduction eliminates the representation
of
the inhomogeneity and
anisotropy of physical media, which have apparent,
experimentally measurable effects (see below).

Numerous additional experimental data have been identified in recent
times
which provide further conditions beyond those of validity of the SR. It
is
important to see that they are all reducible or, equivalently,
expressible via
the arbitrary character of the speed of electromagnetic waves.

Recall that hadrons are not ideal spheres with isolated points in them,
but are instead some of the densest media measured in laboratory until
now. If
spacetime anomalies have been experimentally detected
for media of relatively low density
(such as our atmosphere),
the hypothesis that the SR can be {\it exactly} valid within hadrons
has little scientific credibility. As
one among the numerous arguments, the basic topology of the SR (the
Zeeman
topology of the Minkowskian spacetime) is purely local-differential,
and, as such, it cannot be exactly valid under the notorious {\it
nonlocal}
conditions in the interior of hadrons caused by deep mutual penetration-
overlapping of the {\it wavepackets} of the hadronic constituents (a
feature totally independent from their {\it point-like charges}).

One of the first quantitative studies of the above setting was done by
D. L. Blokhintsev [3a] in 1964, followed by
L. B. Redei [3b], D. Y. Kim [3c] and others.
It should be stressed that the exact validity of the
SR for the {\it center-of-mass behavior} of a
hadron, e.g., in  a particle accelerator is beyond scientific doubt.
Therefore, Refs. [3a-3c]  solely studied the
interior structural problem of hadrons and argued that
a possibility for internal anomalies due to
nonlocal and other effects to manifest themselves in the
outside is given by {\it  deviations from the conventional
Minkowskian behavior of the meanlives
of unstable hadrons with the speed v (or energy E)}.

The reduction of the latter deviations
to arbitrary speeds of light is then consequential.
In fact, the Minkowski metric was originally written by Minkowski [1d]
$\eta = Diag. (1, 1, 1, -c_o^2)$.
Therefore, {\it any deviation} $\hat{\eta}$ {\it from}
$\eta$ {\it necessarily
implies a deviation from} $c_o$, as one can see by
altering any component of the metric and then using Lorentz transforms.

Along these lines, R. M. Santilli [3d] submitted in 1982
the hypothesis that {\it contact-nonpotential interactions in general,
and
those in the interior of hadrons in particular,
 can accelerate
ordinary (positive) masses at speed bigger than
the speed of light in vacuum},
 much along the subsequent
astrophysical measures [2e-2g].
The above hypothesis implies that
{\it  photons travel inside the hyperdense hadrons
at speeds bigger than that in vacuum}. V. de Sabbata and M. Gasperini
[3e] conducted the first phenomenological verification  within
the context of the conventional gauge theories
supporting the hypothesis of Ref. [3d], and actually reaching limit
speeds
up to $75 c_o$ for superheavy hadrons.

The hypothesis of Ref. [3d] is also
supported by the phenomenological calculations
conducted by
H. B. Nielsen and I. Picek [3f] via the spontaneous symmetry breaking
in the Higgs sector of conventional gauge theories, which have resulted
in the anomalous Minkowskian metrics (here written in the notation
above)
\begin{eqnarray}
  \pi: \hat {\eta} = Diag.[(1 + 1.2\times 10^{-3}), (1 + 1.2\times
10^{-3}),
 (1 + 1.2\times 10^{-3}), - c_o^2(1 - 3.79\times 10^{-3})],\\
K:  \hat{\eta} = Diag. [(1 - 2.0\times 10^{-4}), (1 - 2.0\times
10^{-4}),
 (1 - 2.0\times 10^{-4}), - c_o^2(1 + 6.00\times 10^{-4})].
\label{(2,3)}
\end{eqnarray}

As one can see, calculations [3f]
confirm speeds of photons $c = c_o/n > c_o, n < 1$
for the interior of kaons, precisely
as conjectured in Ref. [3d].
Recall that: spacetime anomalies
are expected to increase with the density;
all hadrons have approximately the same size;
and hadrons have densities increasing with mass.
Therefore, results similar to (3) are expected
for all hadrons {\it heavier} than kaons,
 as supported by phenomenological studies
[3e].

The first direct experimental measurements on the behavior of the
meanlife of $K_S^o$ with energy, $\tau (E)$, were done by S. H.
Aronson {\it et al.} [3g] at Fermilab and they
suggested {\it deviations} from the Minkowskian spacetime
in the energy range of
30 to 100 GeV. More recent elaborations [3j,3k] of these measurements
have also confirmed superluminal speeds
inside kaons. Subsequent direct measurements also for $K_S^o$ were
done by S. H. Aronson {\it et al.} [3h]
also at Fermilab, suggesting instead {\it no deviations} of $\tau (E)$
from the Minkowskian form in the {\it different} energy range
of 100 to 400 GeV. Nevertheless, fits [3j,3k] of both data [3g,3h]
have confirmed the superluminal character of the speed, despite
the conventional character for the latter data.

More recently, a test of
the decay law at short decay times was made by the OPAL group
at LEP [3i]. In the latter experiment
the ratio of number of events
$Z^0 \rightarrow \tau^{+} \tau^{-}$
with deviations of  $\tau$ from the conventional law to number of
''normal'' events was $(1.1 \pm 1.4 \pm 3.5)\%$.

In conclusion, the conceptual, epistemological, phenomenological, and
experimental evidence on deviations from the SR within hyperdense media
such
as hadrons or the interior of stars is so overwhelming to render
mandatory
the study of a more adequate theory. In particular, all experimental
evidence suggest a local value of
the maximal causal speed arbitrarily bigger than that in vacuum.

We should stress that all the above anomalies have been and are
hereon strictly and solely referred to
{\it ordinary photons} (those with positive energy) and {\it not}
to tachyons. In fact, tachyons are ruled out as possible candidates
here,
e.g., because the photons for
experiments [2b,2c] are ordinary tardyons before and after the test,
thus preventing the possibility that they acquire a
tachyonic state when passing thru the guides. A similar situation occurs
for all other tests here considered, which all suggest conventional
particles
moving at speeds greater than that of light in vacuum.

The experimental evidence considered herein therefore requires {\it a
new
definition of tachyons}, as the particles traveling
within physical media at speeds bigger than the local
maximal causal speed. This study is not considered here for brevity.

By no means experimental evidence [2,3] exhausts all physical conditions
beyond the SR, and numerous others exist
in the literature. Several of them are somewhat hidden in
{\it ad hoc} parameters which, in reality,
measure the {\it deviation} from the axioms of the SR. This is
the case for the Bose-Einstein correlation where the fit of experimental
data
requires the introduction of the so-called "caoticity parameters" of
unknown physical origin and motivation. A study of this case [4] has
revealed that
these experimental data {\it cannot} be represented via the
strict use of the SR. Therefore, the "caoticity parameters"
provide a {\it bona fide} quantitative representation of the {\it
deviations}
from the SR.

Deviations from the SR are also hidden in theories
which reduce experimental data on the hadronic structure to
{\it hypothetical particles} (namely, particles which cannot be directly
detected on an individual basis), such as the quarks, whose
characteristics
{\it cannot} be defined in our spacetime, and are solely definable in
the mathematical unitary space. It is evident that, under the reduction
of real experimental measurements in our spacetime to hypothetical
particles
outside our spacetime, the characteristics of the latter
can always be adapted to verify the SR.

However, the lack of a possible, direct and independent verifications
of said characteristics of undetectable quarks, combined with the
serious
problematic aspects of quark theories still basically open (such as the
impossibility for a scientific definition
 of gravity for all matter made up of quarks [15c]) and other factors
essentially prevent quark theories from
reaching truly scientific conclusions.

At any rate, studies [5] have shown that Minkowskian anomalies: A) are
fully admitted
by quark theories; B) they merely imply a rescaling of certain
characteristics; and
C) rather than
"destroying" unitary models, they resolve some of the open problems
(such as
confinement), besides permitting otherwise impossible advances (such as
{\it convergent} perturbative expansions for {\it strong}
interactions).
These results then turn quark theories into one of the most compelling
evidence of {\it deviations} from (rather than verification of)
the SR in the interior of hadrons.

 The above entire scientific scene can be represented via the
following historical distinction, which was
well known in the early part of this century and
thereafter ignored:

1) {\it Exterior dynamical problems in vacuum}, in which the SR is here
assumed as valid; and

2) {\it Interior dynamical problems within physical media}, in which
the SR is "inapplicable (and not "violated"
because not constructed
for that scope).

Under the latter perspectives, there is no need for new experimental
data to reach
general physical conclusions. In fact, the conventional
ten conservation laws for exterior
problems imply the exact validity of the fundamental Poincar\'e
symmetry,
with consequential validity of the SR. The same features then imply the
impossibility for the SR to be exactly valid for interior problems
because it
would imply, e.g., that electrons orbit in the core of a star with
conserved angular momentum, and other nonscientific beliefs of the type
of the
perpetual motion.

Equivalently, exterior problems are solely characterized by
"action-at-a-
distance" interactions which, as such, are local-differential,
thus implying the applicability of the mathematical foundations of the
SR.
Moreover, these
interactions are entirely representable via a Lagrangian or a
Hamiltonian,
thus implying the applicability of the physical foundation of the
theory.

By comparison, whether within our atmosphere or within
a star, interior dynamical problems require the additional presence of
{\it contact interactions} as experienced by a space-shuttle during
re-entry
in our atmosphere or,
along much of the same lines, a protons moving in the core of a star.
It is well known that contact interactions can only
occur among {\it extended objects} and have {\it zero range}, thus
implying the
lack of exact applicability, first, of the mathematical foundations
of the SR and,then, of its physical structure.

In conclusion, the validity of the SR for exterior problems and its
inapplicability
for interior problems can be established beyond scientific doubt via the
mere observation (and admission)
of physical reality. Experiments are only needed to establish the
{\it amount of deviations} from the SR for each given interior
conditions.
As a result, any expectation of the necessarily exact validity of the SR
for interior dynamical problems
is outside the boundaries of science, because the only scientifically
open issue
is the selection of a more adequate theory for interior dynamical
problems.

It should be mentioned that, by no means, the above comments exhaust all
criticisms of the SR. An additional serious criticism
is that concerning the {\it ether} to be a "universal substratum", which
is necessary not only
to propagate electromagnetic waves, but also for the existence
of elementary particles (such as the electron)
which are known to be localized oscillation of the
same medium. It is evident that a universal substratum
requires the existence of a universal privileged reference frame
which is strictly prohibited by the SR.

Arbitrary values of the speed of light combined to
the need for a universal substratum and other insufficiencies,
have stimulated the return to
{\it Galilean} forms of relativities for {\it relativistic} conditions.
For brevity, we here indicate Ref. [6] and papers quoted therein.

For completeness we should also mention that, contrary to the majestic
axiomatic consistency of the SR, the {\it general relativity}
GR [7] has remained afflicted throughout this century by serious
problematic
aspects at both classical and quantum levels, when already in its
conventional
formulation for exterior gravitational problems in vacuum. As a matter
of fact, the
unresolved basic problems are so numerous, to discourage even a
partial outline in this note (one may consult the recent article [8a],
Sect. 3 of paper [8b], and the literature quoted therein).

When passing to interior gravitational problems, all limitations of the
SR
carry over in their entirety to the GR, thus implying its lack of exact
character for the conditions considered (because GR is locally
Minkowskian).
The inapplicability begins with the
underlying Riemannian geometry because it is strictly local-differential
and Lagrangian
as compared to the generally nonlocal and nonlagrangian interior
gravitational problems. The inapplicability then extends to numerous
other aspects, from the limited velocity-dependence of the GR (which is
insufficient to represent the notoriously arbitrary velocity-dependence
of interior problems), to the field equations (because of the lack of
full verification
of the Freud identity, lack of torsion, and other aspects).

Finally, we should mention major insufficiencies of both the special and
general
relativities for the {\it classical}  representation of {\it
antiparticles} in
a way truly compatible with their operator counterpart.
\vskip 0.5cm

\noindent {\bf 2. Problematic aspects of existing generalized theories.}
\vskip 0.5cm

\noindent The insufficiencies of
contemporary relativities outlined in Sect. 1 have stimulated the
construction of numerous generalized theories in an attempt to represent
broader
physical conditions Unfortunately, most of them are afflicted by
problems of physical consistency
so serious to prevent their application to experiments.

The problems of physical consistency of broader theories have been
studied in details in memoir [9]. Since they are still vastly unknown to
the general physics audience, an outline appears recommendable here.

One of the reasons for the majestic axiomatic consistency of the SR is
its {\it invariant structure}, that is, a formulation based on a
line element which is invariant under the rotational,
Lorentz and Poincar\'e symmetries. In turn, this invariance implies:

1) {\it The invariance of the fundamental units of space, time, energy,
etc.}
which is an evident pre-requisite for all valid measurements. In fact,
the basic invariant of the theory, the unit of the Poincar\'e symmetry
$I = Diag. ([1, 1, 1], 1)$, represents in a dimensionless way the units
$I = Diag.([1 cm, 1 cm, 1 cm], 1 cm/1 sec)$.

2) {\it The preservation in time of basic notions such as that of
Hermiticity},
which then implies a consistent representation of observables.

3) {\it The uniqueness and invariance of the numerical
predictions};

4) {\it The verification of rigorous causality and probability laws;}

5) {\it the verification of basic axioms of unquestionable consistency};
and other known features all
essential for physical consistency.

The main point of the problems of physical consistency under
consideration here
can be expressed via the following:
\vskip 0.5cm

{\it LEMMA 1: All classical (operator) theories with speed
of light c different than that in vacuum $c_o$
are noncanonically (nonunitarily) related to the special relativity }
\vskip 0.5cm

In fact, by their very conception, the theories here considered are
based on
a structural change of the Minkowski metric which can only be achieved
via a
noncanonical-nonunitary transform
\begin{equation}
\eta \rightarrow \hat {\eta} = U\times \eta \times U^{\dagger},
U\times U^{\dagger} \not = I,
\label{eq:four}\end{equation}
where $\times$ is the conventional associative product.

When formulated on conventional spaces over conventional
fields, the above noncanonical-nonunitary structure of broader theories
has then
disastrous consequences for their physical consistency. In fact, it
implies
(see Ref. [9] for all details and original references):

1') {\it The lack of invariance of the basic units of space, time,
energy, etc.},
as an evident consequence of the very notion of noncanonical-nonunitary
transforms,
with consequential lack of consistent applicability to real
measurements.

2') {\it The lack of preservation in time of Hermiticity and other basic
features},
with consequential absence of physically acceptable observables.

3') {\it The lack of uniqueness and invariance of the numerical
predictions};

4') {\it The general violation of causality and probability laws;}

5') {\it The clear violation of the basic axioms of Einstein's special
relativity}, with consequential rather robust problems of
identifying new covering axioms, proving their mathematical consistency
and then establishing them experimentally; and
other rather serious shortcomings.

In general, the above physical inconsistencies remain hidden in the
existing literature
for various reasons. For instance, a number of theories with arbitrary
speeds of light
are formulated via a seemingly conventional Minkowski space with metric
$\eta = Diag. (1, 1, 1, 1)$ and speeds of light embedded in the fourth
coordinate
$x^4 = ct$. However, the lack of uniqueness of the value c implies
consequential
problematic aspects and ambiguities in the definition of time t.

Generalized theories which exhibit the above
physical shortcomings because possessing a nonunitary time evolution or
for other reasons (e.g., nonlinearity in the wavefunction which implies
the
collapse of Mackay impriimitivity theorem with consequential
invalidity of the SR, or theories with nonassociative envelopes which
'violate Okubo "no quantization theorem")
 are rather numerous. Among them we mention (see Ref. [9] for details):
dissipative nuclear models
with nonhermitean Hamiltonians [17]; statistical models with
external collision terms [18]; nonlinear theories [19]; theories
with nonassociative envelopes [20]; q-deformations [21]; k-deformations
[22];
*-deformations [23]; deformed creation-annihilation algebras [24];
nonunitary statistics [25]; Lie-admissible interior dynamics [26];
noncanonical time theories [27]; supersymmetries [28];
Kac-Moody superalgebras [29] and others.

The reader is encouraged to inspect any of the above generalized
theories,
prove their lack of unitarity on conventional Hilbert spaces over
conventional
fields and verify the occurrence of physical shortcomings 1') to 5').

Other
generalized theories
appear as consistent as the conventional formulation of the SR. The
problematic aspects
emerge in their full light when identifying the necessarily
noncanonical-nonunitary relation to the SR.

This is particularly the case when
Darboux's transforms are used with the reduction of
nonhamiltonian-noncanonical theories in the given coordinate system x of
the observer
to seemingly Hamiltonian-canonical
theories in a mathematical frame x'. However, Darboux's transforms are
not only strictly noncanonical, but above all highly nonlinear, thus
implying the loss of the original inertial character of the reference
frame,
thus implying the evident loss of the SR. At any rate, there is
 the impossibility of conducting actual measurements in the mathematical
frame x'.
Under these conditions, the validity of the SR in the
new mathematical reference frame x' has no physical value.

The problematic aspects of Darboux's and other transforms are the
reason for the insistence in achieving first a
"direct representation", i.e., a representation
in the fixed reference frame of the experimenter, before the
transformation
theory may acquire a physical meaning.

It should be stressed to avoid possible misrepresentation that, by no
means,
{\it all} broader theories verify shortcomings 1') to 5'). As an
illustration,
we indicate the theories worked out by Ahluwalia [3a], Dvoeglazov [10b]
and others
which do have a fully consistent axiomatic structure which bypasses said
shortcomings.

\vskip 0.5cm

\noindent {\bf 3. The novel iso-, geno- and hyper-mathematics and their
isodual.}

\vskip 0.5cm

\noindent Studies conducted by the author during the past decades
have indicated that the primary reason for the physical problematic
aspects
 is
the formulation of generalized theories via the {\it conventional
mathematics
of the SR}, namely, via conventional numbers and fields, conventional
vector and metric spaces, conventional Lie algebras and groups, etc.

To our best knowledge, the occurrence admits no alternatives.
On one side there is the need for noncanonical-nonunitary theories as a
mandatory condition for novelty over existing theories. On the other
side,
when formulated via conventional mathematics, noncanonical-nonunitary
theories
have no physical meaning known to this author because of the
unavoidability
of the problematic aspects of Sect. 2.

The above occurrence left no other choice
than the construction of
a {\it new mathematics}, specifically conceived for the invariant
formulation of
a broader relativity representing arbitrary speeds of light
via noncanonical-nonunitary structures.

After laborious trials and errors, this author proposed
back in 1978 [10] the construction of
a new mathematics with nonsingular well behaved, Hermitean,
generalized, $n\times n$ units
with an unrestricted functional dependence
\begin{equation}
\hat I = \hat I(x, \mu, \omega, ...) = 1/\hat T ,
\label{eq:five}\end{equation}
and then the reformulation of number and fields, vector and metric
spaces,
Lie algebras and groups, etc.
in such a way to admit $\hat I$, rather than I, as the correct left and
right
new unit.

This first requires the lifting of the
 conventional  associative product $A\times B$  among generic
quantities  A,  B into the form $A\hat {\times}  B  = A\times
\hat{T}\times B$
(namely, the lifting of $A\times B$ by
an amount that is the {\it inverse}   of $\hat I$),
under which $\hat I$ is indeed the correct left and right new unit
called the {\it isounit},  while $\hat T$ is
called the  {\it isotopic element}. For consistency,
the {\it totality} of the  of the conventional mathematics must then be
reconstructed,
with no exception known to this author.

One reaches in this way the {\it isonumbers} $\hat n = n\times \hat I$
with {\it isoproduct} $\hat n\hat {\times} \hat m =(n\times m)\times
\hat I$ and
related {\it isofields} $\hat F(\hat n,+,\hat {\times})$; {\it
isospace}s defined over
$\hat F$ with {\it isometric} $\hat {\eta} = \hat T\times {\eta}$;
{\it isolie algebras and groups}; etc.

Since $\hat I$ preserves all the topological properties of I
(nonsingularity,
positive-definiteness, etc.), the  generalized mathematics resulted
to be "axiom-preserving" and was therefore
proposed under the name of {\it isomathematics} [10].

Regrettably, we are not in a position to review the isomathematics for
brevity.
A technical understanding of this paper will however require at least
a rudimentary knowledge of Refs. [11].

For the particular case of theories with arbitrary speeds of light c,
the generalized
unit is given by $\hat I = Diag. (1, 1, 1, c^2)$. Since the unit is the
basic invariant
of any theory, the construction of a theory with new unit $\hat I$
therefore {\it guarantees} the invariance for arbitrary speeds
$c = c_o/n(x, \mu, \omega, ...)$.

It is however evident that isomathematics permits the invariant
formulation of
theories dramatically broader than those with arbitrary speeds of light,
evidently
in view of the arbitrariness of the functional dependence of the
generalized unit.

A yet broader mathematics is the {\it genomathematics} [11], which is
characterized by
{\it nonhermitean} generalized units called {\it genounits}.
In this case we have {\it two} genounits, $I^> = 1/T^>$ and
$^<I = 1/^<T = (I^>)^{\dagger}$, with two corresponding products,
one ordered to the right,
$A>B = A\times T^>\times B$, and
one ordered to the left, $A<B = A\times ^<T\times B$. Genomathematics is
then characterized by two independent liftings of the conventional
mathematics,
one for each of the two ordered products. This includes {\it two}
classes of
genonumbers and genofields, genovector and genohilbert spaces, genolie
algebras and
groups, etc.

A still broader mathematics is the {\it hypermathematics} [11],
which is characterized by nonhermitean {
\it multivalued}   generalized units, $I^> = (I^>_1, I^>_2, I^>_3, ...)$
and $^<I = (^<I_1, ^<I_2, ^<I_3, ...) = (I^>)^{\dagger}$,
with two corresponding multivalued hyperproducts $A>B = A\times
T^>_1\times
B + A\times I^>_2\times B + A\times T^>_3\times B ...$ and
$A<B = A\times ^<T^>_1\times
B + A\times ^<I_2\times B + A\times ^<T_3\times B ...$, and related
hypermathematics.

Once the conventional value of the basic unit +1 dating back to biblical
times is
abandoned in favor of an arbitrary quantity, additional possibilities
emerge
besides iso-, geno- and hyper-mathematics. The additional possibility
significant for this note is the map from
positive into {\it negative} generalized units or, more generally, via
the
map called {\it isoduality} [12b,15]
\begin{equation}
\hat I = \hat I(x, v, \omega, ...) \rightarrow \hat I^d = - \hat
I^{\dagger}
= - \hat I^{\dagger}(- x, -\mu, -\omega,...).
\label{eq:six}\end{equation}
which characterizes yet novel mathematics called {\it isodual
iso-, geno- and hyper-mathematics}.
A particular case is the {\it isodual mathematics}, that is, the
image of the {\it conventional}  mathematics under isoduality with
negative unit -1.

The need for all these generalized mathematics is the following.
Since it is axiom-preserving, {\it isomathematics represents
closed-isolated systems with
Hamiltonian and nonhamiltonian internal effects whose
center-of-mass trajectories are reversible in time},
such as the classical structure of Jupiter
or the operator structure of a hadron with nonhamiltonian internal
effects
when the systems are considered as isolated from the rest of the
universe. In fact,
conventional potential interactions are represented with the
Hamiltonian,
while all nonhamiltonian effects are represented via the isounit. The
reversible character is then ensured by the Hermiticity of the
isounit, $\hat I = \hat I^{\dagger}$, which guarantees the identity of
the
center-of-mass trajectories for motions forward and backward in time.

Despite their generalized structure,
by no means the above systems exhaust physical reality.
{\it Genomathematics is then particularly suited for
the representation of the broader class of open-nonconservative systems
with
Hamiltonian and nonhamiltonian interactions whose center-of-mass
trajectories are irreversible}, such as the classical representation
of the space-shuttle during re-entry in
our atmosphere or a proton moving in the core of a star.
In fact, potential interactions are
again represented with the Hamiltonian and all nonhamiltonian effects
are represented by the
genounits. An axiomatic representation of irreversibility
is then guaranteed by the nonhermiticity of the genounits yielding
two conjugate representations, one for the ordered motion forward in
time with
$I^>$ and $A>B$, and the other for the ordered
motion backward in time with $^<I$ and $A<B$.The conjugation $I^> =
(^<I)^{\dagger}$
then ensures the transition from motion forward to motion backward in
time
under time reversal.

Despite its further generalized character, by no means genomathematics
can represent
the entire physical reality because a number of structures,
such as biological entities, and even
the entire universe are expected to have multivalued structures.
{\it Hypermathematics then permits a representation of multivalued
open-nonconservative systems with Hamiltonian and nonhamiltonian
internal effects
and irreversible structures}, as inherent in the very notion of
multivalued hyperunits and hyperproducts.

Despite its further generalized character, hypermathematics
too cannot represent the entire universe. In fact, it is now known that
conventional, iso-, geno- and hypermathematics cannot provide a
consistent
{\it classical} representation of antimatter which is compatible with
the
established operator counterpart [15].

 Isodual maps are anti-isomorphic like the charge
conjugation, although the latter solely applies at the level
of {\it second quantization},
whole the former apply at all levels {\it beginning}  at the classical
level
and then continuing at the operator level where they
are equivalent to charge conjugation.
Moreover, isodual theories have
{\it negative norm}, thus reversing
the sign of {\it all} physical characteristics of matter,
including charge. In view of these and other reasons,
isodual theories have resulted to be
particularly suited for a {\it classical}  representation of {\it
antimatter}
in a way which is compatible with their operator counterpart [15].

As a result, conventional, iso-, geno- and hyper-mathematics
are used for the representation of {\it matter} in conditions of
increasing
complexity, and, respectively for: point-like abstractions of
particles in exterior conditions
in vacuum; reversible systems of extended particles in interior
dynamical conditions;
irreversible systems of extended particles in interior conditions; and
multivalued irreversible systems of extended particles in interior
conditions.

The isoduals of the conventional, iso-, geno- and hypermathematics are
used for the characterization of {\it antimatter} in corresponding
conditions
of increasing complexity.

For all details, we refer the reader to memoir [11b]
and monograph [11f].
\vskip 0.5cm

\noindent {\bf 4. Isotopic, genotopic and hyperstructural liftings of
the special relativity.}
\vskip 0.5cm

 \noindent The
achievement of an invariant formulation of theories with arbitrary
speeds of light
required the isotopic (that is, axiom-preserving) lifting of the entire
formalism
of the SR, including the Minkowski space, the Poincar\'e symmetry and
the basic
axioms of the SR, whose rudiments were first submitted in Refs. [12].
Operator treatments are available in Refs. [13]. Refs. [14] provides a
partial
list of directly related presentations among a literature
that is rather vast at this writing. For brevity
we can evidently review here only the main lines.

The fundamental isotopy for relativistic theories is the lifting  of the
unit of conventional 4-dimensional theories, $I = diag. (1, 1, 1,
c_o^2)$ of the
Minkowski space and of the Poincare' symmetry,
into a  well  behaved, nowhere singular,
Hermitean and   positive--definite  $4 $-dimensional
matrix $\hat {I} = 1/\hat T$    whose elements   have  an
arbitrary dependence on the needed local quantities.

Since $\hat I = 1/\hat T$ can always be diagonalized, we shall hereon
assume the
form of the isotopic element
$\hat T = Diag. (\hat T_{11}, \hat T_{22}, \hat T_{33}, \hat T_{44}),
\hat T_{\mu\mu} > 0$.

Let $M(x, \eta, R)$ be the  Minkowski space with  spacetime coordinates
$x = \{x^{\mu}\} = \{r, x^{4}\}$, $x^{4} = c_{0}t$, and
 metric $\eta =  Diag. (1,1,1,-c_o^{-2})$ over the reals
$R  =  R(n,+,\times)$.

The original  field $R=R(n,+,\times  )$ is  then lifted  into the
isofield
$\hat{R}=\hat{R}  (\hat{n},+,\hat{\times   })$
 characterized by the above isounit $\hat I$ for  which  all
operations (multiplication, division roots, etc.)  are isotopic.
It  is easy to  see that $\hat{R}$ is locally
isomorphic  to   $R$ by   construction   and,  thus,  the  lifting  $R
\rightarrow  \hat{R}$ is   an  isotopy. Despite   its  simplicity, the
lifting is not  trivial, e.g., because the  notion of primes and other
properties of  number theory depend on the  assumed unit [11a,11g].

Next, we  need  the   lifting  of    the space  $M$   into  the   {\it
isominkowskian     space}
(today also called the {\it Minkowski-Santilli isospace} [14])
$\hat{M}   =      \hat{M}(\hat{x},\hat{\eta
},\hat{R}$) with the same isounit $\hat I$ of the underlying isofield,
which was  first proposed by Santilli in Ref. [12a], and which should
not be confused
with "deformed Minkowski spaces" (because the former is based on a new
unit and related
mathematics, while the latter is based on the conventional
unit and conventional mathematics).

The isominkowski space is characterized by the {\it isocoordinates}
$\hat{x} = x\times  \hat{I}$ on $\hat R$, and the
{\it  isometric} $\hat{\eta  } = \hat T\times \eta$
although,   for  consistency, the latter   should be
defined on $\hat{R}$, thus having the structure $\hat{N} =
(\hat{N}_{\mu\nu}) =
\hat{\eta}\times\hat{I}=(\hat{\eta}_{\mu\nu})\times\hat{I}$.      The
conventional  interval  on  $M$ is then   lifted  into the  {\it
isointerval} on $\hat{M}$ over $\hat{R}$ [12a]
\begin{eqnarray}
\lefteqn{(\hat{x}-\hat{y})^{\hat{2}}=(\hat{x}-\hat{y})^{\mu}
\hat{\times}\hat{N}_{\mu\nu}\hat{\times}(\hat{x}-\hat{y})^{\nu} =
 [(x-y)^{\mu}\times\hat{\eta}_{\mu\nu}\times
(x-y)^{\nu}]\times\hat{I}  =} \nonumber \\
&=& [(x^{1}-y^{1})\times\hat{T}_{11}\times(x^{1}-y^{1})+
(x^{2}-y^{2})\times \hat T_{22}\times(x^{2}-y^{2})+ \nonumber \\
&&+(x^{3}-y^{3})\times \hat T_{33}\times(x^{3}-y^{3})-(x^{4}-y^{4})
\times \hat T_{44}\times(x^{4}-y^{4})\times\hat{I}.
\label{eq:seven}\end{eqnarray}

It  easy to see that $\hat{M}$  is locally isomorphic  to  $M$ and the
lifting $M \rightarrow \hat{M}$ is also an isotopy.
In fact, the lifting of the unit $I \rightarrow \hat I = 1/\hat T$
is compensated by the {\it inverse} lifting of the
metric, $\eta \rightarrow \hat {\eta}
= \hat T\times \eta$. Despite this axiom-preserving character, it is
evident that the isominkowski space is {\it noncanonically-nonunitarily}
related to the conventional space, as necessary for novelty.
For a recent detailed study of the
isominkowskian geometry we refer the reader to
memoir [12g].

The {\it isopoincar\'e symmetry} first submitted  for the first time by
R. M. Santilli in Refs. [12] and it is today called the {\it
Poincar\'e-Santilli
isosymmetry} [15]. In particular, Ref. [12a] presented the isotopies of
the Lorentz symmetry; Ref.s [12b] presented the isotopies of the
rotational symmetry;
Re.s [12c,12d] presented the isotopies of the SU(2)-spin symmetry;
Ref. [12e] presented the isotopies of the Poincar\'e symmetry; Ref.
[12f]
presented the isotopies of the spinorial covering of the Poincar\'e
symmetry; and
Ref. [11g] provided a systematic study of the underlying geometry. For
independent direct treatment of the Poincar\'e-Santilli isosymmetry we
refer
the reader to various studies by Kadeisvili, e.g., Ref. [11c].

The Poincar\'e-Santilli isosymmetry
is the image of the conventional symmetry under isotopies or,
equivalently,
under the condition of admitting $\hat I$, rather than I, as the basic
unit.
As such, the isopoincar\'e symmetry leaves invariant by construction the
generalized
interval (7), thus leaving invariant by construction arbitrary speeds c
for the
particular case $, \hat I = Diag. (1, 1, 1, n^{-2}), c_o = 1$.

Due to the axiom-preserving character, the Poincar\'e-Santilli
isosymmetry
is locally isomorphic to the conventional symmetry. Yet, the two
symmetries
are noncanonically or nonunitarily related, as established by the
different basic units.
Thus the local isomorphic occurs only when each symmetry is treated with
its
own mathematics. Regrettably, we cannot review here for brevity the new
isosymmetry.

Isounit $\hat I = Diag.(1, 1, 1, n^{-2}), c_o = 1$
is not sufficiently symmetrized in spacetime and must be further lifted
into the
expression
$\hat I = Diag. (n_1^{-2}, n_2^{-2}, n_3^{-2}, n_4^{-2}), n = n_4$ which
can be reached
via the application of a conventional (or, more properly, an iso-)
Lorentz transform,
or just the diagonalization of an arbitrary isounit $\hat I$.
In this case $n_4$ preserves
its character as representing the local index of refraction, while the
space $n_k$
can represent space characteristics outside any hope of
representation via the SR, such as the densities
along the three Euclidean axis, extended, nonspherical and deformable
shapes of the
particles considered and other properties of the interior problem at
hand.

The {\it isospecial relativity} (ISR) [12,13] is
the image of the SR under the isotopies.
As such, it is based on the isotopic image of the conventional axioms,
here
expressed for the simpler case $n_1 = n_2 = n_3 = n_s \not = n_4$ (see
Vol. II
of Refs. [13b] for a detailed presentation):
\vskip 0.5cm

{\it ISOPOSTULATE I: The maximal causal speed for interior
dynamical problems is given by }
\begin{equation}
\hat V_{Max} = c_o \times n_s / n_4.
\label{eq:eight}\end{equation}

{\it ISOPOSTULATE II: The addition of speeds for interior dynamical
problems follows the isotopic law}
\begin{equation}
\hat v_{Tot} = ( u + v) / (1 + u_k\times n_k^{-2}.
\times v_k / c_o\times n_4^{-2}\times c_o).
\label{eq:nine}\end{equation}

{\it ISOPOSTULATE III: The dilation of time and the space contraction
for interior dynamical problems follow the isotopic laws}
\begin{equation}
\hat t' = \hat {\gamma}\times t, \hat L ' = \hat {\gamma}^{-1}\times L,
\hat {\gamma} = ( 1 - \hat {\beta}^2)^{-1}, \hat {\beta} =
v_k\times n_k^{-2}
\times v_k / c_o\times n_4^{-2}\times c_o
\label{eq:ten}\end{equation}

{\it ISOPOSTULATE IV: The isodoppler law for
interior dynamical problems is given by the expression (for the
simple case of null aberration)}
\begin{equation}
\hat {\omega}' = \hat {\gamma}^{-1} \times \hat {\omega}.
\label{eq:eleven}\end{equation}

{\it ISOPOSTULATE V: The mass-energy equivalence for interior dynamical
problems
follows the isotopic law}
\begin{equation}
\hat E = m\times c^2 = m\times c_o^2/n_4^2.
\label{eq:twelve}\end{equation}
\vskip 0.5cm

One should note that in the isorelativity the speed of light {\it is
not}, in general,
the maximal causal speed, with the sole
exception of motion in vacuum (where the two speeds trivially
coincide). This is due to the need of avoiding inconsistencies of the SR
for interior conditions such as those
occurring for the Cerenkov light in which the assumption of the speed of
light
as the maximal causal speed in water would imply the violation of the
principle
of causality (because electrons would travel faster). If
causality is somewhat salvaged by assuming that
the speed of light {\it in vacuum}  is
the maximal causal speed {\it in water}
we would have other inconsistencies, such as the violation of the
relativistic law of addition of speeds.
For these and other inconsistencies in
assuming the speed of light as
the maximal causal speed within physical media,
we refer the reader for brevity to monographs [13b].

Note that, the SR and the ISR coincide at the abstract,
realization-free level by conception and construction, as guaranteed by
the
axiom-preserving character of the isotopies. Therefore, the isotopies
resolve problematic aspect 5') of Sect. 2, i.e.,
they avoid the robust
problems of identifying new axioms and proving their axiomatic
consistency.
They merely identify {\it new realizations of the abstract Einsteinian
axioms
of the SR} which, of course, remained top be
verified experimentally (see Sections 5 and 7).
in particular,
criticisms on the axiomatic structure of the ISR
{\it are} criticism on the structure of the conventional SR.

As an incidental note, tachyons should be defined as particles
traveling faster than the local maximal causal speed (8), rather
than the local speed of light c. In fact, only the former exit from
the causal isocone in isospace [12g]. Also, the assumption of tachyons
as
particles traveling faster than the local speed of light would imply
that the
electrons of the Cerenkov light are tachyons which is dramatically
disproved
by physical evidence (e.g., because they can be proved
to be physical particles when exiting water).

The above isotopies of the SR evidently admit sequentially broader
liftings
under genotopies and hyperstructures which yield the
{\it genospecial relativity}  (GSR) and the {\it hyperspecial
relativity}
(HSR), respectively. They are essentially characterized by the
relaxation, first, of the Hermitean and, then, of the
single-valued character of the generalized units.
In particular, these further liftings imply the
relaxation of the totally symmetric character of
line element (7) and then the relaxation of its single-valuedness on
$\hat R$.
These additional features
permit the only invariant formulation of single-valued and
multivalued irreversible systems known to this author
(the communication of other invariant formulations of irreversibility
by the interested reader would be
appreciated). These broader relativities are considerably more involved
on technical grounds and, therefore, they cannot be treated in this
elementary note
(for details, see monographs [13b] and the recent update [13c]).

The {\it isodual isominkowskian spaces, isodual isopoincar\'e symmetry
and
isodual isospecial relativity} [12,13,15] are given by the isodual
images of the preceding formulations
and are used for the representation of {\it antimatter in interior
conditions}.

An important particular case is given by the
{\it isodual Minkowski space, isodual Poincar\'e symmetry and isodual
special relativity} for the
characterization of {\it antimatter in vacuum}. These novel theories
predict
the existence of the {\it isodual photon} [15c,15e] which is
indistinguishable from the ordinary photon for all interactions
except gravitation. If confirmed experimentally, isodual photons
would therefore permit in the future quantitative studies as to whether
far away
galaxies and quasars Are made up of
matter or of antimatter, a study which is strictly impossible with
current theory due to the complete lack in contemporary physics of a
{\it classical} theory of antimatter (let alone the availability
of a consistent theory). For additional properties
on the latter isodual theories we must also refer the reader to Refs.
[12,13,15].

\vskip 0.5cm

\noindent {\bf 5. Direct universality of the new relativities.}
\vskip 0.5cm

 \noindent It is important to see that
{\it the isospecial relativity and its isodual are directly
universal for all possible spacetime
formulations of matter and antimatter possessing a symmetric metric,
respectively}, where "universality" is referred
to all possible spacetime theories of the indicated type, and "direct
universality" is referred to the
representation in the given frame of the observer (without any use of
the transformation theory). This important property can be easily seen
from the
arbitrariness of the isounit or, equivalently, of the isometric and
related
isointerval [7] which includes {\it all} known symmetric spacetime
intervals, including
the Minkowski, Riemann, Finsler other intervals.

The reader should be aware that the above direct universality was
reached only after
proving that {\it the isogeometries underlying the ISR are directly
universality for all
possible (well behaved) nonhamiltonian systems} [11b]. This was
necessary to avoid
the need for a Darboux's transform and related physical problematic
aspects
mentioned earlier.

To illustrate the
direct universality of the ISR, we first note
that Galilean-type spacetime are admitted as a particular case
of the isospecial relativity for isounits given by the tensorial product
of a space and a time component, $\hat I = \hat I_{space}\times \hat
I_{time}$, with
corresponding, rather intriguing (and mostly unexplored until now)
factorization
of the isominkowskian geometry and of the Poincar\'e-Santilli
isosymmetry.

As a second illustration, we note that all the several different
anomalous time evolution laws of Refs. [3] have resulted to be
particular cases of
the isotopic law (10). In fact, as shown by Aringazin [11h], the former
are
obtained from the latter via different expansions
in terms of different parameters and with different truncations [13b].

As a third illustration, it is easy to see that {\it the isominkowski
metric admits as a particular
case all infinitely possible conventional Riemannian metrics}, $\hat
{\eta} = g(x)$.
In fact, all Riemannian metrics are locally Minkowskian,
thus admitting the decomposition
$g(x) = \hat T(x)\times \eta$, where $\hat T$ is a positive-definite
$4\times 4$ matrix, which is precisely the fundamental concept
underlying the isominkowskian geometry.

Therefore, the ISR provides the first known formulation of exterior
gravity
via a Minkowskian-type geometry, with rather important consequences,
such has: the geometric unification of the special and
general relativities; the first formulation of
gravity with a universal symmetry, the Poincar\'e-Santilli
isosymmetry, locally isomorphic to the conventional Poincar\'e symmetry;
the
representation of horizon and singularities via the zeros of the
isotopic unit and isounit; and other advances.

The isominkowskian formulation of gravity also permits the resolution of
some
of the controversies in gravitation that are still open following
about one century of debates, such as: the compatibility of
gravitational and
relativistic conservation laws (which is now resolved via the visual
inspection that the
generators of the related two symmetries coincide); the lack of a
consistent
{\it relativistic} limit of gravitation (which is easily achieved via
the
limit $\hat I\rightarrow I$); and other resolutions [12g].

As a fourth illustration, we indicate that
the arbitrary functional dependence of the isotopic element $\hat T$
in the isometric also permits a direct representation of
interior gravitational problems with the desired arbitrarily nonlinear
dependence on the
velocities, nonlocal-integral effects, and other features of
interior problems outside realistic capabilities by the Riemannian
geometry.

A fifth illustration of the capability of the ISR is that of having
provided
the only known axiomatically consistent grand
unification of gravitation and electroweak
interactions [16] without structural disparities as
in other theories (e.g., electromagnetism
defined on a {\it flat}  spacetime,
with gravitation defined on a {\it curved}  manifold). This result was
precisely due to the isominkowskian representation
of gravity, that is, its formulation in a way axiomatically compatible
with all other interactions.

All the preceding theories are based on a
{\it diagonal spacetime metric}. An additional
class of still broader spacetime theories
is that characterized by {\it nondiagonal isometrics} $\hat {\eta}$ or,
equivalently, {\it nondiagonal isounits $\hat I$} which have
far reaching physical implications, such as the first possibility of
synthesizing neutrons from protons and
electrons as occurring in stars at their first formation [12f].

The geno- and hyper-special relativities are directly universal for all
infinitely possible {\it single- and multi-valued irreversible systems
with nonsymmetric metrics}, respectively, as the reader is encouraged to
verify.
\vskip 0.5cm

\noindent {\bf 6. Simple construction of the new relativities.}
\vskip 0.5cm

\noindent The need for new mathematics has been a major deterrent
for the understanding of the new relativities,
to such an extent that various authors have
preferred noninvariant and physically inconsistent formulations, rather
than entering into the study and use of new mathematics.

It is important to clarify that the use of the new mathematics can be
completely eliminated, and the entire formalism of the
new relativities can be constructed
via truly elementary methods.

First, the ISR camn be simply constructed via the systematic
application of the following
nonunitary transform
\begin{equation}
U\times U^{\dagger} = \hat I ,
\label{eq:thirteen}\end{equation}
to the {\it totality} of the formalism of the SR.

In fact, transform (13) yields the isonumbers $U\times n\times
U^{\dagger} =
n\times \hat I$, the correct form of the isoproduct,
as well as {\it all} conventional operations, special functions and
transforms,
including the correct structure and representation of the Lie-Santilli
isotheory [13a].

Once the structure of the ISR has been achieved in this way, its
invariance is easily
proved via the reformulation of nonunitary transforms in the {\it
isounitary form}
\begin{equation}
U = \hat U\times \hat T^{1/2}, U\times U = \hat U\hat {\times} \hat
U^{\dagger} =
\hat U^{\dagger}\hat {\times} \hat U = \hat I.
\label{eq:fourteen}\end{equation}

The invariance of the ISR then follows. For instance, we have the {\it
numerical invariance of
the basic isounit and related local speed of light} $\hat I\rightarrow
\hat I' = \hat U
\hat {\times}\hat I\hat {\times}\hat U^{\dagger} = \hat I$; the {\it
numerical invariance
of the isotopic element in the isoproduct}; and all other invariances of
the SR as
the reader is encouraged to verify.

Moreover, the above simple method for the construction of the ISR is
also
particularly valuable to generalized existing models in vacuum into
models for
arbitrary speeds of lights. In fact,
one can merely selects the nonunitary transform
$U\times U^{\dagger} = Diag. (1, 1, 1, n_4^{-2})$.

The construction of the GSR is equally elementary, and requires the use,
this time, of {\it two} nonunitary transforms
\begin{equation}
U\times U^{\dagger} \not = I, W\times W^{\dagger} \not = I,
U\times W^{\dagger} = I^>, W\times U^{\dagger} = ^<I,
\label{eq:fifteen}\end{equation}
to the {\it totality} of the formalism of the SR.

In fact, transforms (15) yield the genonumbers $U\times n\times
W^{\dagger} =
n\times I^>$, the correct form of the genoproduct,
as well as {\it all} conventional operations, special functions and
transforms,
including the correct structure and representation of the Lie-Santilli
isotheory [13c].
Once the genotheory is reached in this way, its invariance
is easily proved by rewriting the
nonunitary transforms in their genounitary version.

The HSR can be easily constructed and proved to be invariant
via the mere relaxation of the
single-valued character of the genounits and its outline
is here omitted for brevity [11f]).
\vskip 0.5cm

\noindent {\bf 7. Experimental verifications, applications to other
theories
and concluding remarks.}
\vskip 0.5cm

\noindent Another aspect of the new relativities outlined
in this note that does not appear to have propagated
at this writing in the physics community is the existence of numerous
applications
and experimental verifications among which we quote (for brevity, see
memoir [13a]
for details and all original references): the exact fit of the
experimental data on
behavior of the meanlives of unstable hadrons with speed (thanks to
Isopostulate
III); the exact
fit of the experimental data on the Bose-Einstein correlation for
proton-antiproton annihilation at high and low energies
(thanks to various Isopostulate);
the first exact representation of total nuclear magnetic moments;
the exact representation of the
large differences in cosmological redshifts between quasars and galaxies
when physically related according to
gamma spectroscopy (thanks to Isopostulate IV); the elimination of the
need
for a missing mass in the universe (thanks to Isopostulate V);
and several other experimental verifications.

The iso-, geno- and hyper-special relativities also permit the
invariant formulation of existing theories [17-29] by resolving
their physical inconsistencies as outlined in Sect. 2 (see memoir [9]
for details).

The new relativities and their isoduals are also directly related to a
variety of
broader theories existing in the literature, such as the theories by
Ahluwalia [30a], Dvoeglazov [30b] and others.....

All in all, the above features of the
new relativities are sufficient to warrant further studies.

\vskip 0.5cm

\noindent {\bf Acknowledgments.} The author would like to
thank V. Dvoeglazov for permitting this work to see the light of the day
and for invaluable critical comments.

\end{document}